\documentclass[%
 reprint,
 amsmath,amssymb,
 aps,
showkeys
]{revtex4-2}
\usepackage{graphicx}
\usepackage{dcolumn}
\usepackage{bm}

\usepackage[outdir=./]{epstopdf}
\begin{document}

\preprint{APS/123-QED}

\title{Quantum Simulations of Loop Quantum Gravity}

\author{Swapnil Nitin Shah}
\affiliation{%
Department of Physics, University of Washington, Seattle, WA 98195, USA
}%

\date{\today}

\begin{abstract}
Loop Quantum Gravity (LQG) is one of the leading approaches to unify quantum physics and General Relativity (GR). The Hilbert space of LQG is spanned by spin-networks which describe the local geometry of quantum space-time. Simulation of LQG spin-network states and their dynamics is classically intractable and is widely believed to fall in the Bounded Quantum Polynomial (BQP) time complexity class \cite{LOQC}. There have been many recent attempts to simulate these states using novel and off the shelf quantum computing technologies. In this article, we review three such efforts which utilize superconducting qubits \cite{SC}, linear optical qubits \cite{LOQC} and Nuclear Magnetic Resonance (NMR) qubits \cite{NMR} respectively. The articles chosen for this review represent state of the art in quantum simulations of LQG.       
\end{abstract}

\maketitle
  

\section{\label{Intro}Introduction}
Classical simulations of quantum gravity quickly become intractable as the Hilbert space grows exponentially in the number of degrees of freedom \cite{LOQC,SC}. On the other hand, recent advances in quantum computing technologies have made it possible to simulate such systems (albeit on a small scale) on commercial quantum computers. In this paper, we review state of the art in simulation of LQG spin-networks and evaluation of their transition amplitudes by utilizing various quantum-computing technologies. Before discussing specifics of these implementations, what follows is a brief overview of LQG and its geometric formulation.

Loop quantum gravity is a theory of quantum gravity, which aims to explain the quantum behavior of the gravitational field and more recently, incorporate matter of the Standard Model into the framework established for the pure quantum gravity case. It is an attempt to develop a quantum theory of gravity based on the quantization of Einstein's geometric formulation of General Relativity. This is in contrast to String theory where gravity emerges as one of the vibrational states of the string, namely the graviton which carries the gravitational force. In covariant LQG \cite{CovLQG}, quantum states describe the geometry of a 3-dimensional boundary of a compact 4-dimensional space-time volume. These states are the spin-network states which form the Hilbert space of the theory. The dynamical equation of motion is obtained by a suitable quantization of the Einstein-Hilbert action in GR. This describes the amplitude of evolution from initial to final spin network states or covariantly, the interaction amplitude of spin-network states.  
\subsection{\label{Spin-Networks}Spin Networks}
\begin{figure}[h] 
\includegraphics[width=50mm]{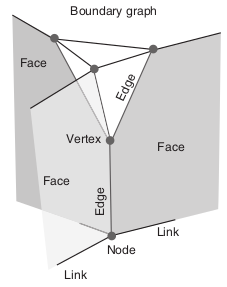}
\caption{\label{fig1}A 2-complex dual to the 4-d triangulation of a space-time volume. (Source: Ref. \onlinecite{CovLQG})}
\end{figure}

In the spin-foam version of LQG \cite{CovLQG}, space-time is discretized by a triangulation $\triangle$ into 4-simplices such that any two neighboring 4-simplices share a single edge (tetrahedron). One considers a 2-complex $\triangle^*$ dual to the triangulation, which connects the 4-simplices of a compact space-time volume (Fig. \ref{fig1}).
The 2-complex induces a directed graph $\Gamma$, dual to a tetrahedral triangulation $\partial\triangle$, on the 3-dimensional boundary of the space-time volume. Each edge $e$ of a face f of the 2-complex is associated with an element $U_e\in$ SL(2,C) which is the holonomy of spin connection over the edge. The triangle of a 4-simplex in $\triangle$ dual to the face f is associated with an element $L_\mathrm{f}$ of the $\mathfrak{sl}$(2,C) Lie algebra. GR (barring the cosmological constant) can be described as a topological BF theory with a linear simplicity constraint on the boundary. In this treatment, the Palatini formulation of Einstein-Hilbert action can be written in the form of BF action with $L_\mathrm{f}$ and $U_\mathrm{f}\equiv\prod_{e\in\mathrm{f}}U_e$ acting as conjugate variables \cite{CovLQG}. On the boundary graph $\Gamma$, these describe the variables $U_l\in$ SU(2) and $L_l\in\mathfrak{su}(2)$ over the links $l$. These are promoted to operators in the quantum theory
\begin{subequations}
\begin{equation}
U_l \;\; \rightarrow \;\; \hat{U}_l \;\;| \;\; \psi(U_l)\equiv\left\langle U_l|\psi\right\rangle
\end{equation}
\begin{equation}
L^i_l \;\; \rightarrow \;\; \hat{L}^i_l \;\;| \;\; \left\langle U_l|\hat{L}^i_l|\psi\right\rangle\equiv-i\kappa\frac{d}{dt}\psi\left(U_l e^{t\tau_i}\right)\bigg|_{t=0}
\end{equation}
\end{subequations}
where $\kappa=8\pi\hbar\gamma G$ and $\tau_i$ are the SU(2) generators. The associated Hilbert space is the kinematical Hilbert space $\mathcal{H}_\Gamma$. The states of $\mathcal{H}_\Gamma$ are not generally SU(2) invariant (local Lorentz invariance of GR mandates SU(2) gauge invariance on the boundary in Palatini formulation). The SU(2) invariant subspace of $\mathcal{H}_\Gamma$ is the physical Hilbert space $\mathcal{H}_\mathrm{phys}$ which is spanned by SU(2) invariant boundary spin network states. In the SU(2) group representation, a general state $\psi\in\mathcal{H}_\mathrm{phys}$ can be described as
\begin{equation}
\psi\left(U_l\right) = \sum_{j_l,k_\mathrm{n}}\mathcal{C}_{j_l,k_\mathrm{n}}\otimes_n i_{k_\mathrm{n}} \otimes_l D^{j_l}\left(U_l\right)
\label{eq_one}
\end{equation}
\noindent
where, $i_{k_\mathrm{n}}$ are 4-valent intertwiners at each node n of the spin-network defined using Wigner-3j symbols as
\begin{equation}
i^m_{k,j}\equiv\sum_{p} (-1)^{(k-p)}
\begin{pmatrix}
j_{l_1}&j_{l_2} &k\\
m_1&m_2&p
\end{pmatrix}
\begin{pmatrix}
k &
j_{l_3} & j_{l_4}\\
-p & m_3 & m_4
\end{pmatrix}
\label{eq_two}
\end{equation}
and $D^{j_l}(U_l)$ is an irreducible representation of $U_l$ labeled by spin $j_l$ associated with a link $l$ of the spin-network. The indices of $D^{j_l}(U_l)$ and $i_{k_\mathrm{n}}$ are appropriately contracted in Eq. (\ref{eq_one}). From Eq. (\ref{eq_two}), a general intertwiner state $\left|\mathcal{I}_\mathrm{n}\right\rangle$ of node n of the boundary graph is described in the spin basis $\left|\otimes_{l_\mathrm{n}}\,j_{l_\mathrm{n}},m_{l_\mathrm{n}}\right\rangle$ as
\begin{equation}
\left\langle\otimes_{l_\mathrm{n}}\,j_{l_\mathrm{n}},m_{l_\mathrm{n}}|\mathcal{I}_\mathrm{n}\right\rangle=\sum_{k_\mathrm{n}}\mathcal{C}_{k_\mathrm{n}}i^{m_{l_\mathrm{n}}}_{k_\mathrm{n},j_{l_\mathrm{n}}} 
\label{eq_intw}
\end{equation}
\noindent
where the sum ranges over all half integers $k_\mathrm{n}$ that satisfy the triangle inequality with all spins $j_{l_\mathrm{n}}$. Indeed, such a state is invariant under SU(2) transformations. Formally,
\begin{equation}
\sum_{l_\mathrm{n}}{\vec{\hat{L}}_{l_\mathrm{n}}} \left| \mathcal{I}_\mathrm{n} \right\rangle = 0
\label{eq_su2_intw} 
\end{equation}
Each such intertwiner is dual to a tetrahedron in the boundary triangulation $\partial\triangle$. From Eq. (\ref{eq_intw}), we find that a spin-1/2 intertwiner state (with spins of all four links $j_{l_n}=1/2$) can be described by a qubit on the Bloch sphere \cite{SC,NMR} (one level corresponding to each $k_\mathrm{n}=0,1$)
\begin{equation}
\left| \mathcal{I} \right\rangle= \cos{\frac{\theta}{2}}\left|0_s\right\rangle+e^{i\phi}\sin{\frac{\theta}{2}}\left|1_s\right\rangle
\label{eq_ten}
\end{equation}
\noindent
where states $\left|0_s\right\rangle$ and $\left|1_s\right\rangle$ are given in the spin basis by
\begin{subequations}
	\begin{equation}
	\left|0_s\right\rangle\equiv\frac{1}{2}\left(\left|0101\right\rangle-\left|0110\right\rangle-\left|1001\right\rangle+\left|1010\right\rangle\right)
	\end{equation}
	\begin{equation}
	\left|1_s\right\rangle\equiv\frac{1}{\sqrt{3}}\left(\left|1100\right\rangle+\left|0011\right\rangle-\left|T_s\right\rangle\right)
	\end{equation}
	\begin{equation}
	\left|T_s\right\rangle\equiv\frac{1}{2}\left(\left|0101\right\rangle+\left|0110\right\rangle+\left|1001\right\rangle+\left|1010\right\rangle\right)
	\end{equation}
\end{subequations}

\subsection{\label{Transition-Amplitudes}Transition Amplitudes}  
In LQG, the transition amplitude of boundary spin network states is given by the Feynman path integral of the discretized classical BF action \cite{CovLQG}
\begin{equation}
W_\triangle(U_l)=\mathcal{N}\int_{\mathrm{SL(2,C)}}dU_e\int dL_\mathrm{f}\,e^{i/8\pi\hbar G\cdot\sum_\mathrm{f}\mathrm{Tr}\left[U_\mathrm{f}L_\mathrm{f}\right]}
\label{eq_three}
\end{equation}
\noindent
where $\mathcal{N}$ is the normalization factor and $G$ is the gravitational constant.
Under the linear simplicity constraint,
Eq. (\ref{eq_three}) can be written as
\begin{equation}
W_\triangle(h_l)=\mathcal{N}\int_{\mathrm{SU(2)}}dh_\mathrm{vf}\prod_\mathrm{f}\delta(h_\mathrm{f})\prod_\mathrm{v}A_\mathrm{v}(h_\mathrm{vf})
\label{eq_four}
\end{equation}
where $h_\mathrm{vf}=g_\mathrm{ev}g_\mathrm{ve'}$ for edges $e,e'\in\mathrm{f}$, a face around the vertex v of the 2-complex such that $g_\mathrm{ve}=g^{-1}_\mathrm{ev}$ and $h_\mathrm{f}=\prod_{\mathrm{v}\in\mathrm{f}}h_\mathrm{vf}$. Also, $U_e=g_\mathrm{ve}g_\mathrm{ev'}$ for an edge $e$ connecting the vertices v and v' of the 2-complex. The vertex amplitude $A_\mathrm{v}(h_\mathrm{vf})$ is given by
\begin{equation}
A_\mathrm{v}(h_\mathrm{vf})=\sum_{j_\mathrm{f}}\int_\mathrm{SL(2,C)}dg'_\mathrm{ve}\prod_\mathrm{f}dj_\mathrm{f}\mathrm{Tr}_{j_\mathrm{f}}\left[Y^\dagger_\gamma gY_\gamma h_\mathrm{vf}\right]
\label{eq_five}
\end{equation}
\noindent
where $g=g_\mathrm{e'v}g_\mathrm{ve}$, $dj_\mathrm{f}=(2j_\mathrm{f}+1)$ and $Y_\gamma$ ($\gamma$ is the Barbero-Immirzi parameter \cite{CovLQG}) is a map from an irreducible representation $D^j_{mn}(U)$ of SU(2) to an irreducible unitary representation $D^{\gamma j,j}_{jm,jn}(U)$ of SL(2,C) (map $Y_\gamma$ satisfies the linear simplicity constraint on the boundary).

\section{\label{Quantum-Simulations}Quantum Simulations}
Two of the central features of covariant LQG are a) Spin-network states describing geometry of a space-like boundary and b) Transition amplitudes describing evolution of these spin-network states.  Recent efforts to simulate LQG focus primarily on these two features. With the limited capabilities of available quantum computing technologies today, most of these simulations are restricted to a single 4-simplex and its boundary states \cite{SC,NMR,PSN}. In this review, we focus on three such articles, each utilizing a different quantum computing technology - superconducting qubits \cite{SC}, linear optical qubits \cite{LOQC} and NMR qubits \cite{NMR} to simulate these objects of import.

\subsection{\label{LQG-SC}LQG with Superconducting Qubits} 
In the article [Ref. \onlinecite{SC}], authors utilize quantum circuits using superconducting qubits to simulate - a) Spin-1/2 intertwiner states (Eq. (6)) and their transition amplitudes and b) Maximally entangled monopole and dipole spin-network states (using 1 and 2 boundary nodes respectively). The authors employed the 5-qubit (Yorktown) and 15-qubit (Melbourne) IBM quantum computers for these simulations. 

\subsubsection{\label{SC-circuits}Quantum Circuit of Spin-1/2 Intertwiner}
\begin{figure}[h] 
	\includegraphics[width=60mm]{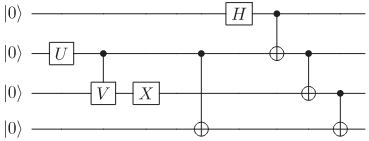}
	\caption{\label{fig2}Quantum circuit to implement a general spin-1/2 intertwiner state (Source: Ref. \onlinecite{SC})}
\end{figure}

In order to implement the spin-1/2 intertwiner, the authors implement a quantum circuit shown in Fig. \ref{fig2}.
In the circuit, the unitary operators $U$ and $V$ are shown below \cite{SC}
\begin{subequations}
\begin{equation}
U=\begin{pmatrix}
c_1 & \sqrt{|c_2|^2+|c_3|^2}
	\\
	\\
\sqrt{|c_2|^2+|c_3|^2} & c^*_1
\end{pmatrix}
\end{equation}
\begin{equation}
V=\begin{pmatrix}
-\frac{c_2}{\sqrt{|c_2|^2+|c_3|^2}} & \frac{c^*_3}{\sqrt{|c_2|^2+|c_3|^2}}
\\
\\
-\frac{c_3}{\sqrt{|c_2|^2+|c_3|^2}} & -\frac{c^*_2}{\sqrt{|c_2|^2+|c_3|^2}}
\end{pmatrix}
\end{equation}		
\end{subequations}
where $c_1, c_2, c_3\in\mathbb{C}$ satisfy
\begin{equation}
\sum c_i = 0 \;\;\; \mathrm{and} \;\;\; \sum |c_i|^2 = 1
\end{equation}
Using this circuit, the authors	implement six representative states of spin-1/2 intertwiners. These are 
$\left|0_s\right\rangle$, $\left|1_s\right\rangle$, $\left|+\right\rangle\equiv\frac{\left|0_s\right\rangle+\left|1_s\right\rangle}{\sqrt{2}}$, $\left|-\right\rangle\equiv\frac{\left|0_s\right\rangle-\left|1_s\right\rangle}{\sqrt{2}}$, $\left|\circlearrowleft\right\rangle\equiv\frac{\left|0_s\right\rangle-i\left|1_s\right\rangle}{\sqrt{2}}$ and $\left|\circlearrowright\right\rangle\equiv\frac{\left|0_s\right\rangle+i\left|1_s\right\rangle}{\sqrt{2}}$ on IBM quantum computers. The circuit is run for 1024 shots for each representative state and the probabilities of measuring a particular spin basis state are computed. 

They use a fidelity measure $F(p,q)$ (Bhattacharya distance) to compare the measured probabilities and theoretical predictions \cite{SC}
\begin{equation}
F(p,q)\equiv\sum_i \sqrt{p_i\,q_i}
\end{equation}
The results are summarized in Table. \ref{table1}. Fidelities for the states are $\approx90\%$ for all the representative states on the 5-qubit system. They sharply fall, however, to $\approx85\%$ on the 15-qubit system, even though the same number of qubits were used in both cases. This reflects the fact that maintaining coherence on larger quantum systems is increasingly difficult.  
\begin{table}[b] 
	\caption{\label{table1}
		Computed values of fidelity for the six representative states on 5-qubit Yorktown and 15-qubit Melbourne (Source: Ref. \onlinecite{SC})
	}
	\begin{ruledtabular}
		\begin{tabular}{lcc}
			\textrm{State}&
			\textrm{Yorktown}&
			\textrm{Melbourne}\\
			\colrule\vspace{-8pt}\\
			$\left|0_s\right\rangle$ & $0.906\pm0.005$ & $0.814\pm0.009$\\
			
			$\left|1_s\right\rangle$ & $0.916\pm0.007$ & $0.856\pm0.008$\\
			
			$\left|+\right\rangle$ & $0.892\pm0.007$ & $0.843\pm0.006$\\
			
			$\left|-\right\rangle$ & $0.915\pm0.007$ & $0.857\pm0.007$\\
			
			$\left|\circlearrowright\right\rangle$ & $0.918\pm0.008$ & $0.856\pm0.008$\\
			
			$\left|\circlearrowleft\right\rangle$ & $0.917\pm0.008$ & $0.851\pm0.007$\\
		\end{tabular}
	\end{ruledtabular}
\end{table}

\begin{figure}[h] 
	\includegraphics[width=46mm]{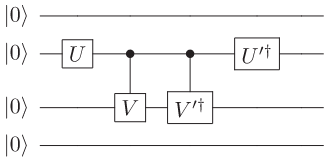}
	\caption{\label{fig3}Quantum circuit to evaluate the transition amplitude between two spin-1/2 intertwiner states (Source: Ref. \onlinecite{SC})}
\end{figure}

In order to simulate the transition amplitudes between two spin-1/2 intertwiner states the authors employ another quantum circuit in Fig. \ref{fig3}. In order to evaluate the amplitude, probability of measuring the state $\left|0000\right\rangle$ is computed after evaluating the circuit for 1024 shots. This scheme is especially useful in evaluating amplitudes of maximally entangled states (considered next) in the intertwiner basis.

\subsubsection{\label{ME-Spin-Networks}Maximally Entangled Spin Networks}
Maximally Entangled Spin-Networks (MESN) are a special class of spin-networks which maximize the entanglement entropy (mutual information) of the state \cite{SC,NMR}. These find applications in Loop Quantum Cosmology, validity of Bekenstein-Hawking blackhole area law, etc. For a given link $l$ of the boundary graph, consider the maximally entangled state 
\begin{equation}
\left|\mathcal{E}_l\right\rangle\equiv\frac{1}{\sqrt{2}}\left(\left|01\right\rangle-\left|10\right\rangle\right)
\end{equation}
One can consider a tensor product of these states over all links of the boundary graph. However, such a tensor product is not generally SU(2) invariant. From Eq. (\ref{eq_one}), it is easy to see that an SU(2) invariant state can be constructed from the tensor product by employing a projector onto the intertwiner states. Such a state is a maximally entangled spin-network state. Formally,
\begin{equation}
\left|\mathrm{MESN}\right\rangle\equiv\hat{P}_G\left|\otimes_l\, \mathcal{E}_l\right\rangle
\end{equation}
where the projector $\hat{P}_G$ in the intertwiner basis is
\begin{equation}
\hat{P}_G\equiv\otimes_\mathrm{n} \left(\left|0_{s,\mathrm{n}}\right\rangle\left\langle 0_{s,\mathrm{n}}\right|+\left|1_{s,\mathrm{n}}\right\rangle\left\langle 1_{s,\mathrm{n}}\right|\right)
\end{equation}
The projection operator $\hat{P}_G$ is not unitary and therefore cannot be implemented as a quantum circuit. However, the components of MESN states in the intertwiner basis can be computed with a circuit by making use of the fact that projector $\hat{P}_G$ does not modify an intertwiner state. Formally, we have from Eq. (6), (14) and (16)
\begin{equation}
\left\langle \mathrm{MESN}\big|\otimes_\mathrm{n}\,\mathcal{I}_\mathrm{n}\right\rangle=\left\langle\otimes_l\,\mathcal{E}_l\Big|\otimes_\mathrm{n}\,\mathcal{I}_\mathrm{n}\right\rangle
\end{equation} 
The authors introduce two maximally entangled spin networks - a) Monopole spin-networks where pairs of links of a single intertwiner are coupled to create a maximally entangled state viz., $(l=2,\mathrm{n}=1)$ and b) Dipole spin-networks which couple links from two intertwiners to create a maximally entangled state viz., $(l=4,\mathrm{n}=2)$. The amplitudes of these states (one each from monopole and dipole networks) in the intertwiner basis are evaluated with circuit equivalents of Eq. (17) on IBM quantum computers. The results obtained are summarized in tables \ref{table2} and \ref{table3}. For the monopole spin-network, the computed results are within $10\%$ of the theoretical predictions on both systems. However, they differ remarkably for the dipole spin-network. This can be attributed to use of double the number of qubits and significant increase in circuit depth with a resultant increase in the accumulated errors \cite{SC}.
\begin{table}[ht] 
	\caption{\label{table2}
		Computed values of amplitude for a monopole spin-network (Source: Ref. \onlinecite{SC})
	}
	\begin{ruledtabular}
		\begin{tabular}{lccc}
			\textrm{Amplitude}&
			\textrm{Theory}&
			\textrm{Melbourne}&
			\textrm{Yorktown}\\
			\colrule\vspace{-8pt}\\
			$\left|\left\langle0_s|\mathcal{E}_{l_1,l_2}\right\rangle\right|^2$ & $0.25$ & $0.23\pm0.01$ & $0.22\pm0.01$\\
			
			$\left|\left\langle1_s|\mathcal{E}_{l_1,l_2}\right\rangle\right|^2$ & $0.75$ & $0.72\pm0.01$ & $0.67\pm0.01$
		\end{tabular}
	\end{ruledtabular}
\end{table}
\begin{table}[ht] 
	\caption{\label{table3}
		Computed values of amplitude for a dipole spin-network (Source: Ref. \onlinecite{SC})
	}
	\begin{ruledtabular}
		\begin{tabular}{lcc}
			\textrm{Amplitude}&
			\textrm{Theory}&
			\textrm{Melbourne}\\
			\colrule\vspace{-8pt}\\
			$\left|\left\langle0_s0_s|\mathcal{E}_{l_1,l_2,l_3,l_4}\right\rangle\right|^2$ & $0.0625$ & $0.008\pm0.002$\\
			
			$\left|\left\langle0_s1_s|\mathcal{E}_{l_1,l_2,l_3,l_4}\right\rangle\right|^2$ & $0$ & $0.003\pm0.002$\\
			
			$\left|\left\langle1_s0_s|\mathcal{E}_{l_1,l_2,l_3,l_4}\right\rangle\right|^2$ & $0$ & $0.009\pm0.002$\\
			
			$\left|\left\langle1_s1_s|\mathcal{E}_{l_1,l_2,l_3,l_4}\right\rangle\right|^2$ & $0.0625$ & $0.008\pm0.003$
		\end{tabular}
	\end{ruledtabular}
\end{table}   

\subsection{\label{LQG-LO}LQG with Linear Optical Qubits}
The previous article dealt with construction of spin-network states and simulation of associated kinematics on actual quantum hardware. In the article [Ref. \onlinecite{LOQC}], authors propose a scalable linear optical framework to evaluate the dynamical transition amplitudes of spin network states on the boundary of a 4-simplex viz., the vertex amplitudes $A_\mathrm{v}(h_l)$ (Eq. (10)). The authors do not physically implement the system but provide evidence from simulations for an illustrative spin-foam vertex amplitude. 

\subsubsection{\label{Penta-SN}Pentagram Spin Networks}
\begin{figure}[h] 
	\includegraphics[width=32mm]{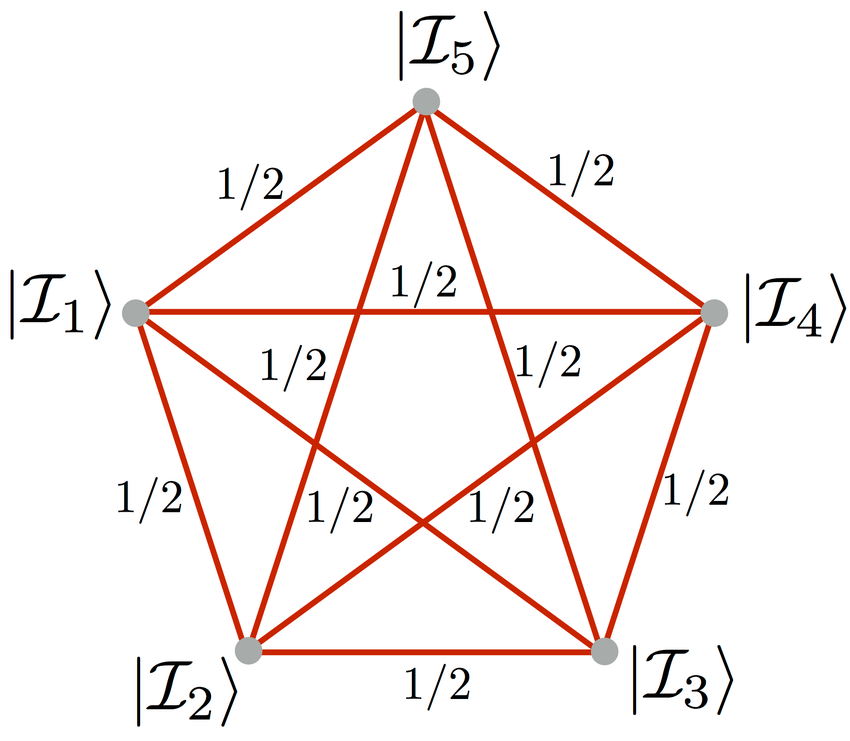}
	\caption{\label{fig5}Pentagram spin-network on the boundary of a 4-simplex with all spins $j_l=1/2$ (Source: Ref. \onlinecite{PSN})}
\end{figure}

\begin{figure}[th] 
	\includegraphics[width=50mm]{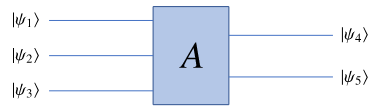}
	\caption{\label{fig4}Spin-foam vertex amplitude as a gate from 3 input to 2 output qubits (Source: Ref. \onlinecite{LOQC})}
\end{figure}

For a single 4-simplex, the boundary graph consists of 5 nodes and 10 links between them \cite{LOQC,PSN}. The associated spin-network is a fully connected \emph{pentagram} spin-network with each node representing an intertwiner (dual to a tetrahedron in $\partial\triangle$) (Fig. \ref{fig5}). 
The components of the vertex amplitude $A_\mathrm{v}(h_l)$ in the intertwiner basis $\left\langle A_\mathrm{v}|\otimes_\mathrm{n}\mathcal{I}_\mathrm{n}\right\rangle$ can be seen as describing the evolution from $m$ intertwiner states to $5-m$ intertwiner states (or covariantly, the interaction between them) \cite{LOQC,NMR}. As earlier, if all $j_{l_\mathrm{n}}=1/2$, the intertwiner states can be described by qubits and the transformation is from $m$ qubits to $5-m$ qubits (Fig. \ref{fig4}) which is evidently non-unitary. Without loss of generality, we will assume $m=3$ henceforth.
For higher order spins $j_{l_\mathrm{n}} > 1/2$, one can always decompose the intertwiner Hilbert space into orthogonal spin-1/2 and spin-0 Hilbert spaces. One can then construct the gate A using simpler gates with inputs and outputs restricted to a subset of these spaces \cite{LOQC}. 

\subsubsection{Spatial Mode Linear Optical Simulator}  
In order to implement the gate A in Fig. \ref{fig4}, the authors extend it to a $12\times12$ unitary operator $U$ given by
\begin{equation}
U\equiv\begin{pmatrix}
A & L\sqrt{I-SS^T}L\\
\\
R\sqrt{I-S^TS}R & -RS^TL
\end{pmatrix}
\end{equation} 
where $L$ and $R$ are unitary and $S$ is the singular value decomposition of $A$ as $A=LSR$. The authors propose using 12 spatial modes of a single photon, out of which 8 are used to encode the input and 4 to encode the output states of gate A. The other 8 input modes are unused and postselection is performed at the output (discard measurements when a photon is not measured in any of the 4 output modes). The article [Ref. \onlinecite{LOQC}] describes how any $\mathrm{N}\times\mathrm{N}$ unitary operator $U$ can be described as a product of identity extended $2\times 2$ unitary matrices $T_{i,j}$ with a diagonal $\mathrm{N}\times\mathrm{N}$ phase correction matrix $D$. Formally,
\begin{equation}
U=D^\dagger T^\dagger_{2,1}\ldots T^\dagger_{N,N-2}T^\dagger_{N,N-1}
\end{equation} 
Such a decomposition admits a straightforward implementation with spatial modes of a photon. Each matrix $T_{i,j}$ only involves mode $i$ and $j$ of the photon and can be easily implemented with a Mach-Zehnder interferometer (MZI). 
\begin{figure}[h] 
	\includegraphics[width=35mm]{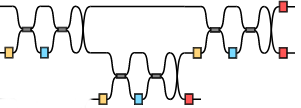}
	\caption{\label{fig6}Spatial mode circuit for 3 photon modes. Yellow boxes - pre-MZI phase, Blue boxes - MZI phase, Red boxes - phase-shifter (Source: Ref. \onlinecite{LOQC})}
\end{figure}    
The phase correction matrix $D$ can be implemented using phase-shifters at each output-mode port. The optical circuit for 3 spatial modes is illustrated in Fig. \ref{fig6}. In this scheme, the total number of MZIs required have a strict upper bound of $(N-1)N/2=66$ for $U$ in Eq. (18), allowing it to be very scalable.

In order to compute $\left|\left\langle\psi_4\psi_5|A|\psi_1\psi_2\psi_3\right\rangle\right|$, a photon is prepared in each of the 8 input modes and probability of the 4 output modes is computed (using postselection). To get the relative phase, a photon is prepared in an equal superposition of two input modes and probability of output modes is computed as earlier. Using this framework, the authors simulate intertwiner-basis components of the vertex amplitude for $j_{l_\mathrm{n}}=2$ and $\gamma=1.2$. To demonstrate that the simulation describes a true quantum process, they evaluate the entanglement entropy of output density matrix for each input mode, which ranges from $0.014 - 0.986$ \cite{LOQC}.

\subsection{\label{LQG-NMR}LQG with NMR Qubits}
In the article [Ref. \onlinecite{NMR}], the authors utilize NMR quantum computing technology to a) Simulate the spin-1/2 quantum tetrahedra (intertwiner states) and b) Compute transition amplitude of pentagram MESN states $(l=10,\mathrm{n}=5)$ (Eq. (17)) using the simulated tetrahedra. They use Crotonic acid molecules with four $^{13}\mathrm{C}$ nuclei (serving as 4 qubits) in a 700 MHz DRX Bruker spectrometer for this purpose.

\subsubsection{\label{NMR-Dihedral}Area and Dihedral Angle}
Area $\mathcal{A}_{l_\mathrm{n}}$ of a triangle (of a boundary tetrahedron) dual to link $l_\mathrm{n}$ of $\Gamma$ is proportional to expectation value of the SU(2) Casimir operator \cite{CovLQG,NMR}

\begin{equation}
\mathcal{A}_{l_\mathrm{n}}\equiv\left\langle \mathcal{I}_\mathrm{n}\bigg|\sqrt{\vec{\hat{L}}_{l_\mathrm{n}}\vec{\hat{L}}_{l_\mathrm{n}}}\bigg|\mathcal{I}_\mathrm{n}\right\rangle=8\pi\hbar\gamma G\sqrt{j_{l_\mathrm{n}}\left(j_{l_\mathrm{n}}+1\right)}
\end{equation}    
Classically, one can choose vectors $\vec{E}_{l_\mathrm{n}}$ normal to the triangle dual to $l_\mathrm{n}$ (face of boundary tetrahedron) such that 
\begin{equation} |\vec{E}_{l_\mathrm{n}}|=\mathcal{A}_{l_\mathrm{n}}
\end{equation}
The dihedral angle $\theta_{km}$ between the faces dual to links $k_\mathrm{n}$ and $m_\mathrm{n}$ is then given by
\begin{equation}
\cos{\theta_{km}}=-\frac{\vec{E}_{k_\mathrm{n}}\cdot\vec{E}_{m_\mathrm{n}}}{\mathcal{A}_{k_\mathrm{n}}\mathcal{A}_{m_\mathrm{n}}}
\end{equation}
A boundary tetrahedron is completely defined by the areas $\mathcal{A}_{l_\mathrm{n}}$ and two dihedral angles $\theta_{km}$ and $\theta_{mk'}$. From Eq. (20) and (22), the dihedral angle operator $\widehat{\cos{\theta_{km}}}$ is	
\begin{equation}
\widehat{\cos{\theta_{km}}}=-\frac{\vec{\hat{L}}_{k_\mathrm{n}}\cdot\vec{\hat{L}}_{m_\mathrm{n}}}{\mathcal{A}_{k_\mathrm{n}}\mathcal{A}_{m_\mathrm{n}}}
\end{equation}

\subsubsection{\label{NMR-Intw}Spin-1/2 Intertwiner with NMR Qubits}
The $^{13}\mathrm{C}$ nuclei of a Crotonic acid molecule form the 4-qubit system to simulate a spin-1/2 intertwiner state (Eq. (6)) which can be individually controlled using a transverse magnetic field \cite{NMR}. The NMR interaction Hamiltonian for this system is given by
\begin{subequations}
\begin{equation}
\hat{H}_\mathrm{int}=\sum_{1\le j\le 4}\pi\nu_j\sigma^j_z + \sum_{1\le j < k\le 4} \frac{\pi}{2}J_{jk} \sigma^j_z \sigma^k_z + \hat{H}_\mathrm{rf}
\end{equation} 
\begin{equation}
\hat{H}_\mathrm{rf}=-\frac{\omega_1}{2}\sum_{1\le j\le 4}e^{-i\left(\omega_0t+\phi\right)}\sigma^j_+ + e^{i\left(\omega_0t+\phi\right)}\sigma^j_-
\end{equation}
\end{subequations}
where $v_j$ is the chemical shift of spin $j$, $J_{jk}$ is the spin-spin coupling, $\omega_1$ parametrizes the magnetic field strength and $\omega_0$ is the oscillation frequency of the applied magnetic field. The 4-qubit system is in the thermal equilibrium state before initialization
\begin{equation}
\rho_{eq}=\frac{1-\epsilon}{16}\mathbb{I}+\epsilon\left(\sigma^1_z+\sigma^2_z+\sigma^3_z+\sigma^4_z\right)
\end{equation}    
where $\epsilon\approx 10^{-5}$. This is transformed into a pseudo-pure state $\rho_{0000}$ using local qubit operations and global gradient-z pulses
\begin{equation}
\rho_{0000}=\frac{1-\epsilon}{16}\mathbb{I}+\epsilon\left|0000\rangle\langle 0000\right| 
\end{equation}
Using transverse magnetic fields for individual qubit rotations and spin-spin interaction for 2-qubit gates, the initial state $\rho_{0000}$ is driven into one of the six intertwiner states $\left|0_s\right\rangle$, $\left|1_s\right\rangle$, $\left|+\right\rangle$,
$\left|-\right\rangle$,
$\left|\circlearrowright\right\rangle$,
$\left|\circlearrowleft\right\rangle$ as well as four other intertwiner states 
$\left|\theta,\phi\right\rangle$,
$\left|\theta,-\phi\right\rangle$, $\left|-\theta,\phi\right\rangle$ and
$\left|-\theta,-\phi\right\rangle$. Here, arguments in the state label indicate coordinates on the Bloch sphere. The dihedral angle operator (Eq. (23)) for spin-1/2 intertwiners is
\begin{equation}
\widehat{\cos{\theta_{km}}}\big|_{j_{l_\mathrm{n}}=1/2}=-\left(\sigma^k_x\sigma^m_x+\sigma^k_y\sigma^m_y+\sigma^k_z\sigma^m_z\right)/6
\end{equation}   
The expectation value $\cos{\theta_{km}}$ is computed as $\mathrm{Tr}\left(\widehat{\cos{\theta_{km}}}\cdot\rho_\mathrm{tetra}\right)$ by using assistive pulses ($\rho_\mathrm{tetra}$ is the simulated intertwiner state) for all ten simulated intertwiners. On comparing with theoretical predictions, authors found fidelity of $\cos{\theta_{km}}$ for their simulated states to be $\approx 95\%$ \cite{NMR}. 

\subsubsection{\label{NMR-AmpSim}Amplitude Simulation of Pentagram MESN}
The authors note that a full simulation of a pentagram MESN state would require a 20-qubit quantum computer which is currently beyond the scope of NMR systems. They perform full quantum tomography of a simulated tetrahedron to determine all components of $\rho_\mathrm{tetra}$. From Eq. (17), magnitude of transition amplitude of MESN states is
\begin{equation}
\left|\left\langle\mathrm{MESN}\Big|\otimes_n \mathcal{I}_n\right\rangle\right|^2 = \left\langle\otimes_l \mathcal{E}_l\Big|\rho_{1234} \otimes \rho_\mathrm{tetra}\Big|\otimes_l \mathcal{E}_l\right\rangle
\end{equation} 
where the authors constrain  $\rho_{1234}=\otimes^{(4)}\left|\circlearrowright\right\rangle\left\langle\circlearrowright\right|$ for simplicity. The complex phase of the transition amplitude is obtained in a similar fashion and the results are summarized in Table. \ref{table4}. The results show that amplitudes computed experimentally are close to theoretical predictions.

\begin{table}[ht] 
	\caption{\label{table4}
		Computed real and imaginary parts of transition amplitude for pentagram MESN states using NMR qubits (Source: Ref. \onlinecite{NMR})
	}
	\begin{ruledtabular}
		\begin{tabular}{lrrrr}
			$\rho_\mathrm{tetra}$&
			\textrm{Re (Th)}&
			\textrm{Im(Th)}&
			\textrm{Re (Exp)}&
			\textrm{Im(Exp)}
			\\
			&$10^{-6}$&$10^{-6}$&$10^{-6}$&$10^{-6}$\\
			\colrule\vspace{-8pt}\\
			$\left|0_s\right\rangle\left\langle 0_s\right|$ & -13.56 & -23.49 & -12.74 & -23.67\\
			
			$\left|1_s\right\rangle\left\langle 1_s\right|$ & 23.49 & -13.56 & 22.16 & -13.16\\
			
			$\left|+\right\rangle\left\langle +\right|$ & 7.02 & -26.20 & 4.32 & -25.62\\
			
			$\left|-\right\rangle\left\langle -\right|$ & -26.20 & -7.02 & -24.59 & -7.98\\
			
			$\left|\circlearrowleft\right\rangle\left\langle \circlearrowleft\right|$ & 0.00 & 0.00 & 0.01 & 0.05\\
			
			$\left|\circlearrowright\right\rangle\left\langle \circlearrowright\right|$ & -27.13 & -46.98 & -25.48 & -44.14\\
			
		\end{tabular}
	\end{ruledtabular}
\end{table}
\noindent
The authors note that the computed amplitude is highest for $\rho_\mathrm{tetra}=\left|\circlearrowright\right\rangle\left\langle \circlearrowright\right|$ which must be true in the classical limit where $\left|\circlearrowright\right\rangle\left\langle \circlearrowright\right|$ represents a regular tetrahedron which glues with four other regular tetrahedra $\rho_{1234}$ to form a 4-simplex \cite{CovLQG,NMR}.

\section{\label{Discussion}Discussion}
In this article, we reviewed three different approaches for LQG simulations of spin-networks and their transition amplitudes. The results obtained in these articles show the unique strengths and limitations of these approaches. Superconducting qubits are the most accessible and controllable quantum computing technology and efficient quantum circuits can be created to simulate spin-1/2 intertwiner states on these platforms with high fidelity \cite{SC}. However, simulating any state beyond monopole spin-networks results in highly unreliable outcomes on these platforms owing to rapid accumulation of errors and decoherence. This necessitates use of quantum error correction and fault tolerant quantum computing. Also, maintaining coherence is increasingly difficult as these systems scale. Photonic qubits with spatial mode encoding provide a viable alternative to superconducting qubits for LQG simulations. The modular scheme of stitching A-gates (one per vertex) to simulate the spin-foam amplitude makes such a system highly scalable and could potentially simulate spin-networks with many nodes \cite{LOQC}. However, designing such a system would require very high precision phase-shifters, MZIs and other optics. Initialization of single photon states in a particular spatial mode would also require a high degree of sophistication. The third approach uses Crotonic acid molecules in an NMR spectrometer to simulate the spin-1/2 intertwiners with very high fidelity \cite{NMR}. The computed values of the transition amplitude for various pentagram MESN states closely match theoretical predictions. The control of individual NMR qubits is extremely difficult in practice, especially if the resonance frequencies of the spins are very close. Also, scaling the system would require molecules with as many controllable nuclear spins as the number of qubits. For a single pentagram spin-network, this would require a molecule with 20 controllable nuclear spins which is beyond the scope of NMR systems today. Owing to this severe limitation, the authors resorted to quantum tomography of each single simulated tetrahedron in order to compute the transition amplitudes. NMR qubit systems are not scalable at this time and therefore do not hold great promise for future LQG simulations beyond a single node.      
   
\section{\label{Conclusion}Conclusion}
The physical Hilbert space of LQG spin networks grows exponentially in the number of nodes. Simulating their dynamics is therefore classically intractable even for small cutoffs for spins $j_l$. With recent advances in quantum computing technologies, various efforts have been made to simulate these complex systems on commercial and novel quantum computing platforms. In this review, we focus on three articles which represent state of the art in LQG simulations, each utilizing a different technology viz., superconducting qubits, optical qubits with spatial mode encoding and NMR qubits. While such efforts have been quite successful at simulating single spin-1/2 intertwiner states, scaling it to spin-networks of practical interest with multiple nodes and higher order spins is still beyond the scope of quantum computing systems of today. Computer simulations with spatial mode encoding of photon states seems promising, however a physical implementation of the proposed framework would be imperative to determining its true scalability. On the other hand, use of error-correction codes and fault tolerant circuits can potentially help scale these simulations on commercial superconducting quantum computers as their qubit volumes increase.    

\nocite{*}
\bibliography{LQG_paper}

\end{document}